# Gender issues in fundamental physics: Strumia's bibliometric analysis fails to account for key confounders and confuses correlation with causation

Philip Ball[1], T. Benjamin Britton[2], Erin Hengel[3], Philip Moriarty[4],

Rachel A. Oliver[5], Gina Rippon[6], Angela Saini[7], and Jessica Wade[8]

[1]Freelance Writer, London

[2]Department of Materials, Imperial College London, London, SW7 2AZ

[3]Department of Economics, University of Liverpool Management School, Liverpool, L69 7ZH (corr. author)

[4]School of Physics & Astronomy, University of Nottingham, Nottingham NG7 2RD, UK (corr. author)

[5]Department of Materials Science, University of Cambridge, 27 Charles Babbage Road, Cambridge CB3 0FS

[6]Professor Emeritus of Cognitive NeuroImaging, Brain Centre, Aston University, Birmingham B4 7ET

[7]Science Journalist, London

[8]Department. of Physics, Imperial College London South Kensington Campus, London SW7 2AZ

Alessandro Strumia recently published a survey of gender differences in publications and citations in high-energy physics (HEP). In addition to providing full access to the data, code, and methodology, Strumia (2020) systematically describes and accounts for gender differences in HEP citation networks. His analysis points both to ongoing difficulties in attracting women to high-energy physics and an encouraging—though slow—trend in improvement.

Unfortunately, however, the time and effort Strumia (2020) devoted to collating and quantifying the data are not matched by a similar rigour in interpreting the results. To support his conclusions, he selectively cites available literature and fails to adequately adjust for a range of confounding factors. For example, his analyses do not consider how unobserved factors—*e.g.*, a tendency to overcite well-known authors—drive a wedge between quality and citations and correlate with author gender. He also fails to take into account many structural and non-structural factors—including, but not limited to, direct discrimination and the expectations women form (and actions they take) in response to it—that undoubtedly lead to gender differences in productivity.

We therefore believe that a number of Strumia's conclusions are not supported by his analysis. Indeed, we re-analyse a subsample of solo-authored papers from his data, adjusting for year and journal of publication, authors' research age and their lifetime "fame". Our re-analysis suggests that female-authored papers are actually cited more than male-authored papers. This finding is inconsistent with the "greater male variability" hypothesis Strumia (2020) proposes to explain many of his results.

In the conclusion to his paper, Strumia states that "*..dealing with complex systems, any simple interpretation can easily be incomplete…*". We agree entirely. Strumia's simple—and, more importantly, simplistic—analysis and interpretation are far from complete.

## Biased literature review

Strumia (2020) notes that there is a "vast literature" dealing with gender differences in STEM subjects. Scientific analyses of gender differences should represent this literature in an even-handed and unbiased manner; as Del Giudice et al. (2019) highlight, "*An honest, sophisticated public debate on sex differences demands a broad perspective with an appreciation for nuance and full engagement with all sides of the question*." That appreciation for nuance and full engagement is not present in Strumia (2020). For example:

- On p. 2, Strumia asserts that "*No significant biases have been found in examined real grant evaluations [Ceci et al.(2014), Marsh et al.(2011), Ley et al.(2008), Mutz et al.(2012)] and referee reports of journals [Borsuk et al.(2009), Ceci et al.(2014), Edwards et al.(2018)]*". Yet a large body of literature—which he fails to cite—reaches the opposite conclusion. (See, for example, Burns et al., 2019; Card et al., 2020; Dworkin et al., 2020; Fox and Paine, 2019; Helmer et al., 2017; RSC, 2019; Steinberg et al., 2018; Witteman et al., 2019 and references therein.) Strumia (2020)'s lack of balance in citing relevant work is misleading and arguably disingenuous. An objective analysis of gender differences should aim to be neither.

- Strumia (2020) also notes that theoretical modelling of citations is "*affected by questionable systematic issues*". We assume the use of "questionable" here is meant to capture relationships that are difficult to identify and quantify—*e.g.*, due to a lack of suitable controls. We agree entirely that all efforts to study citations must be mindful of these limitations. Again, however, Strumia selectively cites just one paper that "*attempts to control for some social factors*", namely Caplar et al. (2017). Other studies analysing more restricted samples have come to the opposite conclusion (see, e.g., Card et al., 2020; Hengel and Moon, 2020).

We do not, of course, expect Strumia (2020) to review the entire research on gender differences in STEM. We believe, however, that a fairer representation of the literature is warranted, especially considering the contentious nature of the topic.

## Confounders and statistical reanalysis

In addition to selectively citing the literature, Strumia (2020) fails to consider the potential impact of a broad range of confounding factors on gender differences in citations and the publication process. To help highlight the problems this introduces, one of the authors (EH) re-analysed a subset of Strumia's data.

The re-analysed data contain 5,599 solo-authored articles—5,386 authored by a man and 213 authored by a woman—published in five high profile physics journals from 2010 to 2016 (inclusive). The selection criteria were designed to address the influence of key confounders in Strumia's data. First, we restricted the sample to solo-authored articles to account for the fact that male physicists are more likely to be senior authors on papers involving much larger research teams.[1] Second, restricting the data to articles published in a small set of well-known journals made it easier to confirm the quality of gender assignment in Strumia's data.[2] By including journal-year fixed effects, we also better account for differing citation patterns between fields.[3] Third, younger articles have had less time to accrue citations and older articles are disproportionately by male authors. For that reason, we also restrict our analysis to newer articles—*i.e.*, those published between 2010 and 2016—as well as controlling for journal-year fixed effects.

---

[1] In contrast, Strumia (2020), Hengel (2017) and Hengel and Moon (2020) assume each co-author on a paper contributes equally to it. This relationship, however, is unlikely to be linear when there are a very large number of co-authors as there often are in fields such as high-energy experimental physics.

An additional difference between Strumia's study and our own is that we analyse data at the article-level instead of aggregating citations over authors. This allows us to better address the following issues discussed in Hengel and Moon (2020):

1) Male authors disproportionately cluster at the very top and very bottom of the citation distribution, but raw citation counts are censored below at zero and unbounded from above. This generates a non-linear mapping from quality onto citations that depends on the former's variance. When used as a proxy for quality, average citations for male-authored papers will, as a result, generally place too much weight on high-citation papers and not enough weight on low-citation papers compared to the average for female-authored papers.

    To deal with this issue, we transform raw citation counts with the inverse hyperbolic sine function (asinh). We stress, however, that our results do not meaningfully change if we use raw citation counts as the dependent variable, instead.

2) Unobserved (or uncontrolled for) confounders boost citations conditional on quality and disproportionately correlate with articles and authors located in the distribution's right tail— *e.g.*, winning a prestigious award. A related concern is that the citations a paper accumulates aren't fixed in time. As a result, they could be influenced by the future success or failure of a paper's authors—*i.e.*, even among non-superstar physicists, a stronger publishing record later on probably drives citations to earlier work, all else equal (for evidence see, *e.g.*, Bjarnason and Sigfusdottir (2002)). Both factors potentially correlate with gender: for example, women produce fewer papers than men and are proportionately less likely to win the Nobel prize.

    To deal with these issues, we control for the year in which an author was first published and her total number of lifetime publications.[4] Our results should therefore be interpreted as gender differences in citations between authors who began their careers around the same time and had accumulated similar lifetime "fame" at the time citations were collected.

Assuming citations are not perfectly explained by these variables,[5] gender differences present in Strumia's complete, unadjusted data should also be present in the conditioned, restricted data if male physicists are, in fact, biologically more productive and/or produce higher quality work than women.[6] In other words, if male talent is more variable, then male physicists' work will be (on average) higher "quality" (as defined by citation count), all else equal. As a result, a paper by a famous female physicist published in 2010 in Physical Review D should be (on average) cited less than a paper by a similarly famous male physicist that was also published in Physical Review D in 2010.

---

[2] We verified the gender of all women and men with fewer than two citations to their papers in the restricted dataset. We found 21–26 percent of people classified as women were actually men. Moreover, their solo-authored articles tended to receive a disproportionately low number of citations.

[3] Strumia (2020), in contrast, adjusts for field by normalising a paper's citation count by the length of its own reference list, which roughly correlates with field (Strumia, 2020 Equation (1)). (We obtain similar results using his normalised indicator of citations.)

[4] To account for age, Strumia (2020) weights each author by one-half times the inverse of the proportion of authors of his same gender who first published in the same year he did (Equation (2)). For example, if 300 authors—two of whom were female—first published in 1995, then each female author would be weighted by 75 whereas each male author would be weighted by roughly 0.5.

[5] Effectively, the distribution of citations cannot collapse to a degenerate distribution after conditioning on these variables. As evidence against this possibility, it *does not* appear that citations are homogenous, conditional on, *e.g.*, journal.

This is not what we observe. Our evidence suggests female-authored papers receive about 12 log points *more* citations than male-authored papers, conditional on covariates. This figure is weakly statistically significant.

Fig. 1 below shows the distribution of citations among male- vs. female-authored papers in this sample for both transformed and non-transformed citations. Note that the distribution of citations to male-authored papers closely overlaps with the distribution of citations to female-authored papers across the entire range of the distribution.

Estimated gender differences in citations at the mean for each of the five journals are shown in Fig. 2. They consistently suggest either no statistically significant gender gap in citations or a citation gap that favours women.

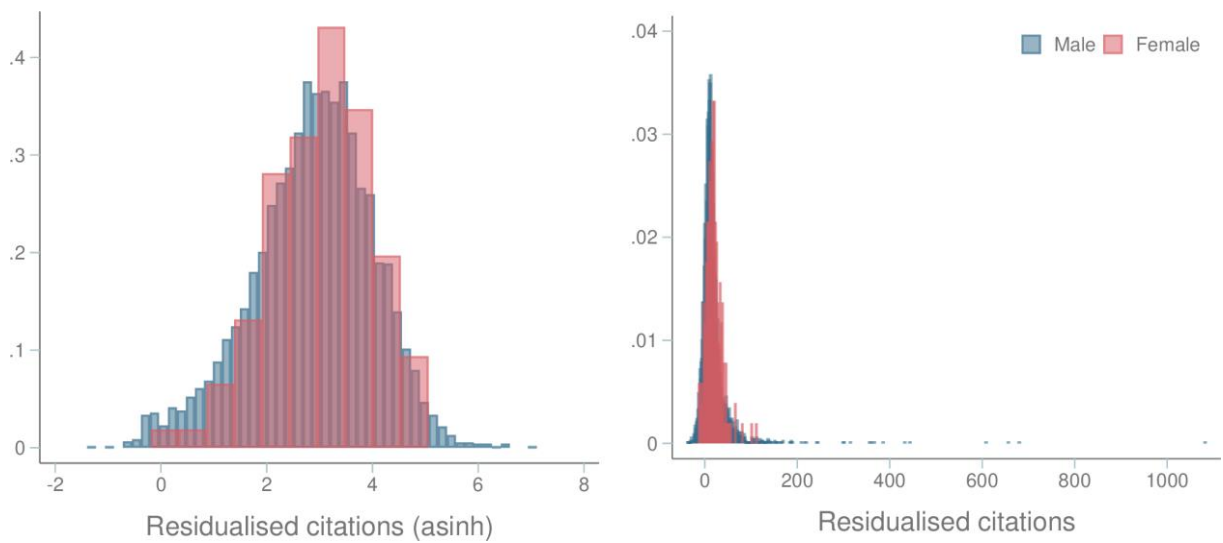

*Figure 1 Distribution of citations for solo-authored papers*

*Note.* Graphs display histograms of asinh transformed (left) and raw (right) citations for solo-authored papers by men (blue) and women (pink) published between 2010–2016 (inclusive) in *Physical Review D*, *Astrophysical Journal*, *Journal of High Energy Physics*, *Physical Review Letters* and *Physics Letters B*. Citations have been residualised with respect to year-journal fixed effects, fixed effects for each author's year of first publication and total lifetime number of publications. Data from Strumia (2020).

---

[6] Note that gender differences in variability are equivalent to gender differences in (conditional) averages. Presumably, all physicists are drawn from the top half of the distribution of "talent". Thus, greater variability in men implies that average male talent is higher than average female talent, conditional on being a physicist.

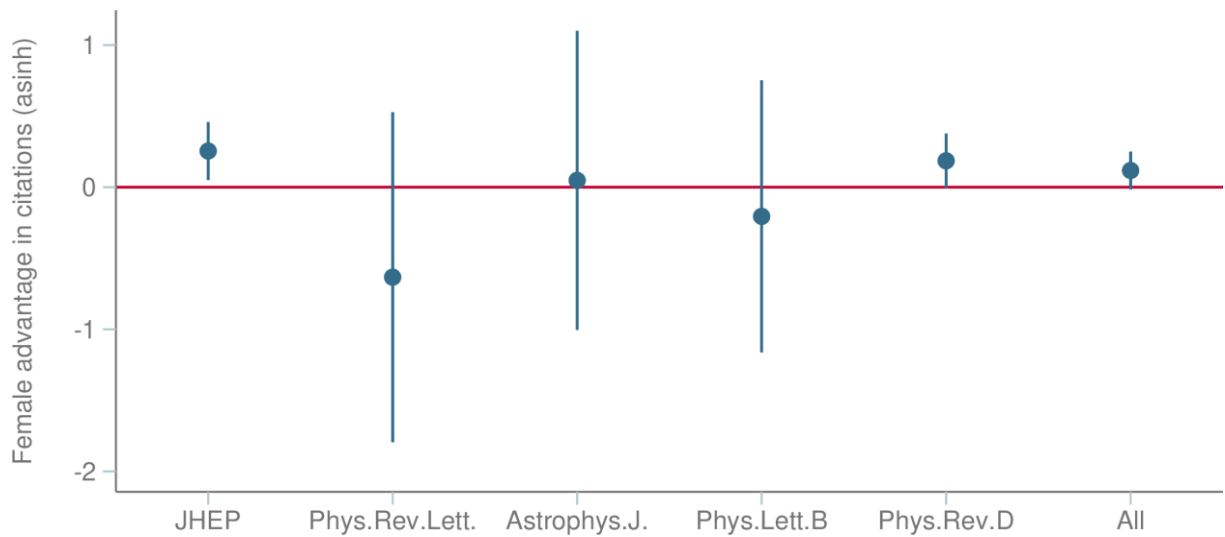

*Figure 2 Gender differences in citations across journals*

*Note.* First five figures display conditional mean gender differences in citations (asinh); a positive value indicates female-authored papers are cited more than male-authored papers, conditional on included controls. They are estimated by regressing citations (asinh) on a dummy variable equal to one if the author was female in the sub-samples of solo-authored articles published between 2010–2016 (inclusive) in *Journal of High Energy Physics* (JHEP), *Physical Review Letters* (Phys.Rev.Lett), *Astrophysical Journal* (Astrophysical Journal) and *Physical Review D* (Phys.Rev.D). Final figure is the estimated gender difference in the pooled data. Lines represent 95% confidence intervals from standard errors clustered on an author's year of first publication. All estimates control for fixed effects for year or year-journal interactions, year of first publication and total lifetime number of publications. Data from Strumia (2020).

## The higher male variability hypothesis

Despite the admission in Strumia (2020) that the simple interpretation laid out therein "can easily be incomplete", the data are nonetheless explained in the context of the highly contentious higher male variability (HMV) hypothesis and a biological basis of difference, together with gendered differences in interests. Once again, there is a lack of appropriate representation and citation of the relevant extensive literature base.

The HMV hypothesis is widely contested and debated; both Gray et al. (2019) and Stevens and Haidt (2017) provide systematic and even-handed discussions. Note, in particular, the geographical variation highlighted by Gray et al. (2019) in relation to the HMV hypothesis, which counters the claims that any observed gender differences are biological in origin:

> *"... we find that there is significant heterogeneity between countries, and that much of this can be quantified using variables applicable across these assessments (such as test, year, male-female effect size, mean country size, and Global Gender Gap Indicators)."*

Geographical and temporal heterogeneity are consistently observed in a variety of measures of gender disparity in STEM (see, for example, Breda et al., 2018; Kane and Mertz, 2012; Nollenberger et al., 2016). Counter-intuitively, however, the so-called "gender equality paradox" put forward by Stoet and Geary (2018), and cited on a number of occasions in Strumia (2020), is the claim that countries with a higher level of gender equality tend to have less gender balance in STEM fields. We note that Stoet and Geary's arguments have been undermined significantly by the many deficiencies in their data analysis highlighted by Richardson et al. (2020) (including those that have necessitated the publication of a corrected version of Stoet and Geary (2018)).

Strumia's undue and unwarranted confidence in the HMV interpretation—given his admission that the data are influenced by "questionable systematic issues"—is such that the abstract closes with the claim that the "quantitative shape" of the data can be "fitted by higher male variability". As highlighted by Hyde (2014),

> "...even if there is slightly greater male variability for some cognitive measures, this finding is simply a description of the phenomenon. It does not address the causes of greater male variability, which could be due to biological factors, sociocultural factors, or both."

The sociocultural factors to which Hyde refers are exceptionally difficult to model (and correct for) yet play an integral role, as discussed by Kalender et al. (2019) in their study of gendered patterns in the construction of physics identity.

Strumia (2020) also does not provide any direct evidence for the causal link he suggests between the HMV hypothesis, biological determinism, and citation rates. Instead, there is nothing more than an inference based on drawing a comparison with what are termed "psychometric observations". This is an entirely unjustified conflation of correlation and causation and has no place in a rigorous interpretation of the data. Only properly controlled studies allow for a robust distinction between correlation and causation—a fundamental tenet of all statistical analysis. Strumia (2020) admits that those controlled studies are simply not possible.

Similarly, the homogeneity of the people/things dimension (Spelke, 2005; Thelwall et al., 2019) is very much overstated in Strumia (2020) in support of the argument that a lack of interest underpins the level of participation by women in HEP. Thelwall et al. (2019), whose work is cited by Strumia in the context of the people-things metric, devote a great deal of time in the conclusion of their paper highlighting the deficiencies in their approach:

> "Thus, the people/things dimensions can only provide a partial explanation for gender differences in topic choices across the full spectrum of academia because there are many important exceptions...Given that the current research has not attempted to assess any cause and effect relationships, deviations from the people/thing dimensions could also be due to other factors within academia that deflect people from pursuing their interests, such as editorial, departmental or funding policies."

None of this important yet difficult-to-quantify nuance is captured by the discussion in Strumia (2020).

Another confounding factor is vocal criticism of women within academia by individuals such as Strumia, who may well be contributing to the problem of women feeling unwelcome in physics. As Halpern et al. (2007), whose paper is cited in Strumia (2020), put it,

> "We conclude that early experience, biological factors, educational policy, and cultural context affect the number of women and men who pursue advanced study in science and math and that these effects add and interact in complex ways. There are no single or simple answers to the complex questions about sex differences in science and mathematics."

The issues outlined above demonstrate that extreme care must be taken when arguing for causal relationships among variables in a system where confounders are exceptionally difficult to deal with (or, indeed, to identify in the first place). A full, appropriately controlled statistical analysis of the cumulative influence of bias—both explicit and implicit, including differing levels of scrutiny in the publication process (Hengel, 2017), discrimination, harassment, bullying, parental and domestic responsibilities, access to research funding, and variations in teaching, committee, and pastoral care workloads on the work of women in HEP (and hence on their outputs and citation rates)—would be needed to interpret Strumia's observations in a scientifically robust and objective manner.

## Bibliometrics as a proxy for scientific quality

There is yet a broader issue surrounding Strumia's methodology: the analysis implicitly assumes not only a direct link between citation rates (normalised as per Equation (1) of Strumia (2020)) and scientific quality, but, as noted above, infers—with no direct quantitative evidence—a further link to the HMV hypothesis. Moreover, it is assumed that all authors contribute equally to a paper. Strumia recognises, to some extent, that this assumption of equal contributions is problematic – "*…there is no warranty that each author contributed to each paper*"—but his justification is very far from compelling: "*Despite this, data show that the total fractionally-counted bibliometric output of collaborations scales, on average, as their number of authors [Rossi et al (2019)], suggesting that large collaborations form when is scientifically needed and that gift authorship does not play a role.*" A lack of gift authorship, *i.e.*, the inclusion of an author who contributed little or nothing at all, is a far cry from the extreme assumption of equal contributions across the entire authorship of a paper. (This is another justification for the restriction of our analysis in the previous section to solo-authored publications.)

In dismissing the role of sociological factors, Strumia (2020) also subjectively (and rather inconsistently) appeals to the reader's perception of the prestige of top authors: "*A physicist might read their names and conclude that no sociological confounder can wash away most of them*". MacRoberts and MacRoberts (2010) describe the process of biased citing which involves a number of considerations including the "halo effect" (exemplified by the preceding quote from Strumia (2020)), in-house citations, obliteration, and the Matthew effect. Each of these effects, and others (Cowley, 2015)) can lead to disproportionate citations of the primary literature. There is also very good evidence that seminal papers can often be cited but not actually read (Ball, 2002; Simkin and Roychowdhury, 2003).

More broadly, the use of citations as a measure of scientific quality is highly questionable and has been the subject of significant debate for decades. (See Leydesdorff et al., 2016 for a review.) We note that Strumia highlights that "*traditional metrics (such as citation counts, h-index, paper counts) now fail to provide reasonable proxies for scientific merit in fundamental physics*". His solution—the introduction of what he terms an "individual citation" metric, *i.e.*, Equation (1) of Strumia (2020), is not a compelling strategy to isolate scientific quality from citation numbers. As Asknes et al. (2019) highlight,

> "*Research quality is a multidimensional concept, where plausibility/soundness, originality, scientific value, and societal value commonly are perceived as key characteristics…citations reflect aspects related to scientific impact and relevance, although with important limitations… there is no evidence that citations reflect other key dimensions of research quality*"

Strumia's reasoning regarding the efficacy of citation metrics as a measure of scientific quality is also notably circular: "*…bibliometric indicators, which can be used as reliable proxies for scientific merit, being significantly correlated to human evaluations such as scientific prizes.*" The attempt to draw a distinction here between bibliometric indicators and "human evaluations such as scientific prizes" is telling. Bibliometrics are not a wholly objective quantitative proxy for scientific merit—they are just as much a human evaluation as is a scientific prize. The award of scientific prizes very often involves detailed consideration of citation rates, and the prestige that stems from a scientific prize will in turn drive more citations for a particular scientist. It is naïve to imagine that bibliometrics are an independent measure of scientific merit whose usage is somehow justified by their correlation with the award of "human evaluations such as scientific prizes."

## Conclusions

The analysis presented in Strumia (2020) suffers from a number of deficiencies that severely undermine the inferences drawn and the conclusions reached therein. In particular, the contributions of a wide variety of potential confounding factors are not only unaddressed but are dismissed with no justification. This unwarranted dismissal, coupled with a fundamental confusion of correlation and causation when drawing inferences, is especially problematic in a study that claims to provide an unbiased assessment of the role of gender in citation patterns. Instead, the analysis throughout is very far from neutral or disinterested; the data are interpreted within an ideologically-motivated context— as is clear from both the paper itself and previous work by the same author, namely Strumia (2018) and Strumia (2019).

In that broader context, Strumia's analysis is substantively problematic. The views he espouses have had an impact not only across the physics and STEM communities but have been widely reported across the international media (including Conradi, 2019; Giuffrida and Busby, 2018; Nicholson, 2018; Young, 2019). Physics as a discipline is broadly acknowledged to struggle with both the recruitment and retention of women (Eddy and Brownell, 2016; Kalender et al., 2019; Porter and Ivie, 2019). The widespread dissemination of derogatory, and unsubstantiated, views about women's ability in this sphere is not going to overcome this problem. It will be off-putting to talented women who might be considering a career in the sciences and contributes to the hostile environment endured by many women (and other minorities) currently working in physics. There is an increasing body of research that suggests that an individual's performance in an academic context can be harmed by an awareness that others' perception of their work might be distorted by stereotypes (see, for example, Casad et al., 2017; Kalender et al., 2019; Shapiro and Williams, 2012). The associated cultural implications can result in minoritised individuals not contributing or disengaging from their academic communities.

Ultimately, the work presented in Strumia (2020) is not merely a flawed, biased, and ideologically motivated analysis. It is also likely to be actively harmful to the progress of women in physics, at a detriment not only to many individuals but to our entire community.

# Funding

TBB acknowledges funding of his Research Fellowship from the Royal Academy of Engineering.# Data and replication files

Data and replication files for all analyses presented in this paper are available on GitHub (https://github.com/erinhengel/strumia-qss).

# References

Aksnes DW, Langfeldt L and Wouters P (2019) Citations, citation indicators, and research quality: an overview of basic concepts and theories. *SAGE Open* 9(1): 1–17.